\documentclass{aa}
\usepackage{epsfig}
\usepackage{natbib}
\bibpunct{(}{)}{;}{}{,}

\newif\ifAMStwofonts
\def\spose#1{\hbox to 0pt{#1\hss}}
\def\simlt{\mathrel{\spose{\lower 3pt\hbox{$\mathchar"218$}}
     \raise 2.0pt\hbox{$\mathchar"13C$}}}
\def\simgt{\mathrel{\spose{\lower 3pt\hbox{$\mathchar"218$}}
     \raise 2.0pt\hbox{$\mathchar"13E$}}}

\def\hi{H\,{\sc i}~}
\def\hii{H\,{\sc ii}~}

\begin{document}
%\thesaurus{02.08.1; 09.01.1; 11.05.2}
\title{Dynamical and chemical evolution of NGC1569}
\author{Simone Recchi\inst{1}\thanks{recchi@astro.univie.ac.at} 
        \and Gerhard Hensler\inst{1} 
        \and Luca Angeretti\inst{2}
        \and Francesca Matteucci\inst{3}}
\offprints{S. Recchi}
\institute{
Institute of Astronomy, Vienna University, T\"urkenschanzstrasse 17, 
A-1180 Vienna, Austria\and
INAF - Osservatorio Astronomico di Bologna, Via Ranzani 1, 40127
Bologna, Italy\and
Dipartimento di Astronomia, Universit\`a di Trieste, Via G.B. Tiepolo, 11,
34131 Trieste, Italy}
\date{Received  /  Accepted   }

\abstract{ Blue Compact Dwarf and Dwarf Irregular galaxies are
  generally believed to be unevolved objects, due to their blue
  colors, compact appearance and large gas fractions.  Many of these
  objects show an ongoing intense burst of star formation or have
  experienced it in the recent past.  By means of 2-D hydrodynamical
  simulations, coupled with detailed chemical yields originating from
  SNeII, SNeIa, and intermediate-mass stars, we study the dynamical and
  chemical evolution of model galaxies with structural parameters
  similar to NGC1569, a prototypical starburst galaxy.  A burst of
  star formation with short duration is not able to account for the
  chemical and morphological properties of this galaxy.  The best way
  to reproduce the chemical composition of this object is by assuming
  long-lasting episodes of star formation and a more recent burst,
  separated from the previous episodes by a short quiescent period.
  The last burst of star formation, in most of the explored cases,
  does not affect the chemical composition of the galaxy, since the
  enriched gas produced by young stars is in a too hot phase to be
  detectable with the optical spectroscopy.  Models assuming the
  infall of a big cloud towards the center of the galaxy reproduce the
  chemical composition of the NGC1569, but the pressure exercised by
  the cloud hampers the expansion of the galactic wind, at variance
  with what observed in NGC1569.

}
\maketitle
\bigskip\bigskip

\section{Introduction}
\label{sec:intro}

Among dwarf galaxies, Blue Compact Dwarfs (BCDs) and Dwarf Irregulars
(dIrrs) are characterized by large gas content and often active
star formation (SF).  They also show very blue colors and low
metallicities and are therefore commonly believed to be systems
similar to primeval galaxies.  They are consequently ideal targets to
study the feedback between star formation and interstellar medium.
They have also been suggested to be the local counterparts of faint
blue objects (Babul \& Rees 1992; Lilly et al. 1995; Ellis 1997).

It has recently become clear that most of these objects show the
presence of at least intermediate age stars (Kunth \& \"Ostlin 2000),
but their contribution to the global metallicity and energy budget of
the galaxy is still unknown.  It is interesting to simulate galaxies
whose light and colors are dominated by young stars and to see whether
their chemical and morphological properties can be explained by a
recent burst of star formation or whether older episodes of SF are
required.

In general, the SF in BCDs is described as a {\it bursting} process
(Searle, Sargent \& Bagnuolo 1973), namely, short, intense episodes of
SF are separated by long inactive periods.  A {\it gasping} mode of SF
(long episodes of SF of moderate intensity separated by short
quiescent periods) is instead often used to describe the star
formation in dIrrs (Aparicio \& Gallart 1995).  Galaxy candidates
experiencing gasping SF are for instance NGC6822 (Marconi et al.
1995), Sextans B (Tosi et al. 1991), and the LMC (Gallagher et al.
1996).  These two different SF modes have been studied in the
framework of chemical evolution models (Bradamante, Matteucci \&
D'Ercole 1998; Chiappini, Romano \& Matteucci 2003a; Romano, Tosi \&
Matteucci 2005a), and they produce similar results, because the
chemical enrichment mostly depends on the gas consumption.  It is
therefore not easy to discriminate between these two different SF
scenarios on the basis of chemical evolution models alone.  Moreover,
the chemical evolution of galaxies changes under conditions imposed by
dynamical processes.  Therefore, a chemo-dynamical approach is needed.

In the context of a long term project aimed at studying the dynamical
and chemical evolution of dwarf galaxies, in this paper we present the
results on NGC1569, a gas-rich, metal-poor and possibly young dwarf
galaxy.  In the following, we briefly summarize the main results
presented in a series of papers (Recchi et al. 2001; 2002; 2004) where
we have analyzed the dynamical and chemical evolution of IZw18, the
most metal-poor galaxy locally known. Although these models were
tailored on this specific object, they give a hint on the enrichment
of a galaxy in the first hundreds of Myrs of evolution, and in
addition, IZw18 will become useful in our discussion.  Thus, the main
results are:

\begin{itemize}
  
\item A galactic wind develops in almost all the models.  This wind
  carries away mostly the metals freshly produced during the burst.
  
\item SNeIa are important triggerers of the galactic wind and the
  metals produced by this kind of SNe (mainly iron-peak elements) are
  ejected more easily than the products of SNeII (mostly
  $\alpha$-elements).
  
\item In models with a bursting SF, the freshly produced metals cool
  down in a short time-scale (of the order of a few 10$^7$ yr). This
  can justify the so-called ``instantaneous mixing'' approximation,
  commonly assumed in chemical evolution models (Matteucci 1996) and
  apparently confirmed by recent observations of the most metal-poor
  stars in the Galactic halo (Spite et al. 2005; Arnone et al. 2005).
  In models with a gasping SF instead, the metals produced by the last
  burst of SF mix with the surrounding ISM in a much longer
  time-scale.
  
\item Either by assuming a gasping SF scenario or a bursting one, {\it
    no stars older than 300--500 Myr} are required in order to
  reproduce the characteristics of IZw18.

\end{itemize}

In summary, a significant contribution of intermediate-mass stars is
decisive in the chemical and dynamical evolution of IZw18, but no old
stars are required.  From a chemodynamical point of view, IZw18 is
therefore a young galaxy.  Recently, Izotov \& Thuan (2004) showed
that, although their data go 1-2 mag deeper than the tip of the Red
Giant Branch phase, no stars are detected in this stage of their
evolution.  This sets a robust constraint on the age of this galaxy
(of the order of a few hundred Myrs) and lets them conclude that {\it
  IZw18 is a bona fide young galaxy} (at least young in a cosmological
sense), in agreement with the chemodynamical calculations and at
variance with recent claims that most of the Blue Compact Dwarf
galaxies (including IZw18) host a very old (with an age of several
Gyrs) population of stars (e.g. \"Ostlin 2000).  It is however worth
pointing out that new analyses of the Izotov \& Thuan (2004) ACS data
(Momany et al. 2005; Tosi private communication) show probably the
presence of RGB stars in IZw18, therefore this object might be older
than assumed by Izotov \& Thuan.

The subject of this paper, NGC1569, is often classified as a DIrr,
whereas IZw18 is a BCD galaxy.  Despite the different classification,
these two objects show similar properties: both are in the aftermath
of an intense burst of SF, are very metal poor and have an extremely
large gas content.  NGC1569 is located near the Local Group.  Its
distance is 2.2 Mpc (Israel 1988) but different values can be found in
literature.  It is also characterized by a significant extinction
(E(B-V) = 0.56; Israel 1988), due to its low galactic latitude.  Its
stellar population is dominated by three super star clusters (De
Marchi et al. 1997; Origlia et al. 2001), but a large number of star
clusters of smaller dimensions has been detected (Anders et al.
2004).  The total dynamical mass of NGC1569 is M$_{\rm dyn}$ $\sim$
3.3 $\times$ 10$^8$ M$_\odot$ (Israel 1988), one third of which is
neutral hydrogen (Stil \& Israel 2002).

The oxygen abundance of NGC1569 is 12 + log (O/H) = 8.19 $\pm$ 0.02
(Kobulnicky \& Skillman 1997).  If oxygen is assumed to be a tracer of
the global metallicity, it implies a global metallicity of $\sim$ 0.3
Z$_\odot$ (assuming the Asplund et al. 2005 solar chemical
composition; a value of $\sim$ 0.2 Z$_\odot$ is obtained adopting the
Anders \& Grevesse 1989 one).  The N/O ratio lies $\sim$ 0.2 dex above
the average log (N/O) detected in the most metal-poor galaxies
($\overline{{\rm log (N/O)}} \sim$ -1.6; Izotov \& Thuan 1999).
According to Kobulnicky \& Skillman (1997), log (N/O) = -1.39 $\pm$
0.05.  These observations revealed also a global uniformity of the
chemical composition of NGC1569, indicating either an exceptionally
fast mixing of the freshly produced metals or that the youngest
generations of stars are not contributing to the global metallicity
budget of this galaxy.

Outflowing gas from NGC1569 has been inferred from the kinematics of
an extended system of H$\alpha$ filaments (Heckman et al. 1995; Martin
1998) and from a diffuse X-ray emission (Heckman et al. 1995; Martin,
Kobulnicky \& Heckman 2002; Ott, Walter \& Brinks 2005). Martin et al.
(2002) attempted to evaluate the chemical composition of the hot
medium, obtaining metallicities larger than those in the \hii regions
(but see also Ott et al. 2005, in which a good fit is obtained when
assuming metallicities of the hot gas similar to the ones measured in
the \hii regions).  Finally, there is observational evidence of the
presence of extended \hi clouds and complexes surrounding NGC1569
(Stil \& Israel 1998).  In particular, there is a series of gas clumps
in the southern halo probably connected with a \hi arm present in the
western side of the galaxy.  These \hi features can be attributed to
the debris of a tidally disrupted big cloud infalling towards NGC1569
(M\"uhle et al.  2005).

The aim of this paper is to study the dynamical and chemical evolution
of NGC 1569, by performing 2-D hydrodynamical simulations in
cylindrical coordinates, coupled with detailed chemical yields coming
from SNeII, SNeIa, and winds from intermediate-mass stars (IMS). In
this context, we study the case of a single, dominating burst of star
formation and of a more complex history of star formation and to
investigate which scenario better reproduces the chemical composition
of this object.  We will also analyze the chemical composition of
different gas phases and we will finally study the impact of the
infall of cold \hi clouds on the chemical and energetic budget of the
galaxy.  In Sect. 2 we summarize our knowledge about the stellar
component and the past star formation history of NGC1569.  In Sect. 3
we briefly summarize the model and the adopted assumptions in the
simulations.  Results are presented in Sect. 4 and, finally, some
conclusions are drawn in Sect. 5.

\section{The star formation history of NGC1569}
\label{sec:SF}

The stellar component of NGC1569 is dominated by two super star
clusters, usually called 'A' and 'B' (nomenclature of Arp \& Sandage
1985) although SSC--A has found to be a double system (De Marchi et
al. 1997).  The two complexes SSC--A and SSC--B are located $\sim$ 8''
($\sim$ 80 pc assuming a distance of 2.2 Mpc) apart.  Their mass is
estimated to be (5--16) $\times$ 10$^5$ M$_\odot$ and their age of the
order of a few tens Myrs (Anders et al.  2004).  Although young stars
dominate the light of NGC1569, Hubble Space Telescope (HST)
observations revealed the presence of older stars.  Vallenari \&
Bomans (1996) found that NGC1569 experienced a recent episode of SF,
lasting about 100 Myr and terminating 4 Myr ago.  They also found
hints for an older and weaker episode of SF, from 1.5 Gyr until 150
Myr ago.  Greggio et al. (1998) found a SF constant over the last
$\sim$ 100 Myr with little or no quiescent periods.  The SF rate they
infer was of the order of 0.5--3 M$_\odot$ yr$^{-1}$, depending on the
adopted IMF slope.  Martin et al.  (2002), by fitting the H$\alpha$
luminosity, found a much weaker SF rate of 0.16 M$_\odot$ yr$^{-1}$
for the last episode of SF.  Aloisi et al.  (2001) found an age
gradient in NGC1569, in the sense that the youngest stars are mostly
concentrated in the 3 super star clusters, intermediate-age stars are
almost uniformly distributed, whereas oldest stars are mostly located
at the outskirts of the starbursting regions.

Recently Angeretti et al. (2005, hereafter A05), by analyzing HST's
NICMOS/NIC2 data in a field of 200 $\times$ 200 pc$^2$ (encompassing
the 3 largest star clusters), concluded that, in order to reproduce
the Color-Magnitude Diagrams (CMDs) of NGC1569 with synthetic ones,
assuming a Salpeter IMF between 0.1 and 120 M$_\odot$, the best choice
of the SF history is:

\begin{itemize}
  
\item A most recent episode of SF, from 37 to 13 Myr ago, at a rate of
  0.13 M$_\odot$ yr$^{-1}$ (equivalent to 3.2 M$_\odot$ yr$^{-1}$
  kpc$^{-2}$).
  
\item An intermediate episode of SF, commencing 150 Myr ago and
  finishing 40 Myr ago (therefore separated from the recent episode of
  SF by a very short quiescent period of 3 Myr), at a rate of 0.04
  M$_\odot$ yr$^{-1}$ (equivalent to 1 M$_\odot$ yr$^{-1}$
  kpc$^{-2}$).
  
\item An older episode of SF, ending 300 Myr ago (therefore implying
  150 Myr of inactivity between this episode and the intermediate one)
  and commencing either 600 Myr ago (at a rate of 0.05 M$_\odot$
  yr$^{-1}$, equivalently 1.2 M$_\odot$ yr$^{-1}$ kpc$^{-2}$), or 2
  Gyr ago (at a rate of 0.01 M$_\odot$ yr$^{-1}$, equivalently 0.25
  M$_\odot$ yr$^{-1}$ kpc$^{-2}$)

\end{itemize}

It is worth pointing out that A05 cannot exclude a low SFR during the
inter-burst phases and in the last 13 Gyr. In addition, the history of
SF derived in the NICMOS area studied by A05 could differ from that of
the external parts mainly because of the stochastic origin of the SF.

\section{The model}
\label{sec:model}

\subsection{The gravitational potential and the initial gas 
configuration}
\label{subsec: potwell}

We simulate a galaxy model resembling NGC1569 by means of a 2-D hydro
code in cylindrical coordinates.  It is worth pointing out that
NGC1569 is characterized by a pronounced arm of gas in the western
side and by an extensive system of H$\alpha$ filaments, therefore it
is definitely non-axisymmetric.  A 2-D geometry can treat the global
properties of the galaxy but not the complex \hi features in detail
(see also Sect. 4.7).  The ISM is initially isothermal and in
hydrostatic equilibrium with the galactic potential well and the
centrifugal force.  The galactic potential well is given by a
quasi-isothermal spherical dark halo and an oblate stellar component.
The dark halo has a core radius of 1 kpc and is truncated at a radius
of 20 kpc.  Its total mass is 1.4 $\times$ 10$^9$ M$_\odot$ but only a
very small fraction of this dark matter ($\sim$ 5\%) is within the
optical size of NGC1569.  The estimates of the total mass by Reakes
(1980) point towards a lower value ($\sim$ 5 $\times$ 10$^8$
M$_\odot$) but, as pointed out by Martin (1998), the relative
contribution of dark and baryonic matter, at least in the central
region of the galaxy, is poorly constrained.
  
As already described in Recchi et al. (2001), in order to reproduce a
gas distribution with the correct ratio between minor and major axis
($\sim$ 0.5; Reakes 1980), we introduce a stellar component described
by an oblate King profile.  Due to the relatively low mass, compared
with the total mass of the galaxy, the contribution of the newly born
stars to the total potential well is neglected and therefore the
potential is fixed in time.  The details on how to build a rotating
gas distribution in equilibrium with a given potential well can be
found in D'Ercole \& Brighenti (1999).

\subsection{Chemistry and supernovae}
\label{subsec:chemistry}

We couple our hydrodynamical simulations with detailed chemical yields
originating from SNeII, SNeIa, and winds from IMS.  The SNeIa rate is
calculated according to the so-called Single-Degenerate scenario
(namely C-O white dwarfs in binary systems that explode after reaching
the Chandrasekhar mass because of mass transfer from a red giant
companion).  It is worth noticing that, in the case of episodes of SF
of short duration, a significant number of SNeIa explodes after a
timescale of the order of 10$^8$ yr (Matteucci \& Recchi 2001),
therefore they can alter the chemical composition of the galaxy in a
timescale much shorter than often invoked for this kind of SNe (of the
order of 1 Gyr).  In the so-called Double-Degenerate scenario
(explosion after merging of two white dwarfs as a result of angular
momentum losses via gravitational wave radiation) one has to take into
account the delay due to the emission of gravitational waves.
However, in the case of instantaneous bursts of star formation, also
the Double-Degenerate scenario predicts a peak in the SNeIa rate after
100--200 Myr (see e.g. Ruiz-Lapuente \& Canal 1998).

In order to mimic continuous episodes of SF, we calculate the mass of
stars formed between $t$ and $t$ + $\Delta t$, with $\Delta t$ $=$
10$^5$ yr.  This {\it Stellar Population} (SP) is treated as a
starburst.  We calculate the evolution, in space and time, of 8
chemical elements of particular astrophysical relevance, namely H, He,
C, N, O, Mg, Si, Fe.  The mass return rate from each of these chemical
elements is calculated according to eq. (1) of Recchi et al. (2004),
namely we sum up the contributions of all the SPs which can provide
enrichment of the element $l$ at the evolutionary time $t$.  Any of
these chemical element must obey a conservation law of this kind:

\begin{equation}
{\partial \rho^l \over \partial t} + \nabla \cdot ({\rho^l \bf{v}}) 
= \dot\rho^l,
\end{equation}
\noindent
where $\rho^l$ represents the mass density of the $l$-th element,
$\bf{v}$ the fluid velocity and $\dot \rho^l$ is the mass return rate
due to the contribution of SNeII, SNeIa and winds from IMS.

For massive and intermediate-mass stars we adopt the yields of Meynet
\& Maeder (2002) (hereafter MM02) since they seem to better reproduce
the chemical composition of IZw18 (Recchi et al. 2004).  It is however
worth noticing that the MM02 calculations do not take into
consideration later phases of the stellar evolution (in particular the
third dredge-up and the hot-bottom burning phase), and may
underestimate the production of primary elements, in particular
nitrogen.  Although the 1-D simulations of Herwig (2004) predict a
very efficient third dredge-up, Marigo (2003) showed that, assuming
variable molecular opacities, the efficiency of the hot-bottom burning
can significantly decrease, in particular for the more massive
asymptotic giant branch stars.  The yields of MM02, at variance with
previous models, are obtained from self-consistent complete
simulations of the stellar evolution and most of the primary nitrogen
production stems from the physical effect of rotation.  The importance
of these new implications on the chemical evolution of galaxies has
not yet been fully tested.  Moreover, MM02 models are the ones which
better reproduce the log (N/O) and log (N/Fe) ratios in the solar
neighbourhood, as well as the solar abundance of helium (Chiappini,
Matteucci \& Meynet 2003b) and they also help to explain the
relatively high log (N/O) and log (N/Fe) recently found in very
metal-poor stars (Chiappini, Matteucci \& Ballero 2005).  A discussion
on the effect of a different choice of stellar yields can be found in
Sect. 4.1.

We adopt the MM02 model with Z=0.004 and $v_{ini}$=300 km s$^{-1}.  $
This metallicity is the one which better reproduces the CMD of NGC1569
in the work of A05.  For consistency reasons we adopt it as the
metallicity of the stellar population.  A05 used the Padova tracks
(Fagotto et al.  1994) which are, in the low metallicity range, only
calculated for Z=0.004 and Z=0.0004.  Therefore A05 can just rule out
a metallicity as low as Z=0.0004.  We cannot therefore exclude a
metallicity slightly lower than Z=0.004, in particular for old stellar
populations.  Therefore the nitrogen and carbon production from old
stars should be considered as an upper limit.  This compensates, at
least qualitatively, the fact that, since MM02 do not follow the whole
evolution of the star, the final nitrogen production might be larger
than tabulated from these authors.  The IMF slope of NGC1569 is well
constrained by the mass/luminosity ratio and is close to the Salpeter
slope (Sternberg 1998).  We therefore adopt a single-slope Salpeter
IMF between 0.1 and 60 M$_\odot$.

We assume, consistently with Recchi et al. (2004) a low thermalization
efficiency for SNeII and a much larger value for SNeIa.  This
assumption is justified by the fact that stars are always born in
molecular clouds, irrespective of the SF history of the galaxy.  The
first exploding SNe (the SNeII) have to get rid of cold and dense
material before contributing to the total ISM energy budget, therefore
they can loose a significant fraction of the mechanical energy
produced by the explosion, at variance with what happens for SNeIa.
In our simulations the SF is supposed to take place in an area of 200
$\times$ 200 pc$^2$ (the field investigated by A05), therefore both
energy sources (SNeII, SNeIa and winds from IMS) and mass return are
concentrated inside this region.  Greggio et al. (1998) performed
observations with the WFPC2 onboard HST (where each WF covers a region
of 80" $\times$ 80" and the PC covers an area of 35" $\times$ 35") and
found that, outside the planetary camera, there are a limited amount
of stars. In fact, the density of the detected stars in the PC is
about 30 times higher than in each WF.  Therefore, most of the SF in
NGC1569 has occurred in the central region and assuming an area of 200
$\times$ 200 pc$^2$ for the input of energy and mass is a rather safe
assumption, although A05 cannot exclude weak episodes of SF outside
the investigated area.

The model contains also a metallicity-dependent cooling function
(according to the tabulated values of B\"ohringer \& Hensler 1989) and
the thermal conduction.  The central resolution of the grid is 5 pc;
the grid size expands logarithmically with a size ratio between
adjacent zones of $\sim$ 1.03.

\subsection{Star formation and gas mass}
\label{subsec:sf}

As described in Sect. 2, the stellar population in NGC1569 is
dominated by stars younger than a few tens Myrs (Anders et al. 2004),
although the presence of older stars has been inferred (Vallenari \&
Bomans 1996; Greggio et al. 1998).

We therefore adopted two possible SF histories for NGC1569.  The first
is a single burst of star formation, lasting for 25 Myr.  The SF rate
calculated by Greggio et al. (1998) for a Salpeter IMF is 0.5
M$_\odot$ yr$^{-1}$.  A weaker SF rate for the present burst (0.13
M$_\odot$ yr$^{-1}$) has been found in A05.  In this set of models,
given the uncertainties of this value, we keep the SF rate as a free
parameter.  The duration of this SF episode, according to Anders et
al. (2004) has been assumed to be 25 Myr.  We have assumed three
possible SF rates for these models: one with 0.1 M$_\odot$ yr$^{-1}$
(model NGC -- 1), one with 0.3 M$_\odot$ yr$^{-1}$ (model NGC -- 2)
and one with 0.5 M$_\odot$ yr$^{-1}$ (model NGC -- 3).  In all models,
the initial gas content of the galaxy is 10$^8$ M$_\odot$.  For the
model NGC -- 1, each SP has therefore a mass of 10$^4$ M$_\odot$.
Also for this relatively small SP, it is reasonable to assume a
Salpeter IMF between 0.1 and 60 M$_\odot$ and neglect statistical
fluctuations.  If we adopt the formulation of Weidner \& Kroupa (2004)
and a theoretical upper stellar mass of 150 M$_\odot$, the derived
upper stellar mass cut-off for a star cluster of 10$^4$ M$_\odot$ is
$\sim$ 100 M$_\odot$ (see Weidner \& Kroupa 2004, their fig.  5).

\begin{table*}[ht]
\caption{Parameters for the NGC1569 models}
\label{model}
\begin{centering}
\begin{tabular}{ccccc}
  \hline\hline
\noalign{\smallskip}

  Model  &  SF episodes & SF rate (M$_\odot$ yr$^{-1}$) 
& SF duration (Myr) & M$_{gas}$ (M$_\odot$) \\
\noalign{\smallskip}

  \hline 
  NGC -- 1 & 1 & 0.1              &  25 &                         10$^8$ \\
  NGC -- 2 & 1 & 0.3              &  25 &                         10$^8$ \\
  NGC -- 3 & 1 & 0.5              &  25 &                         10$^8$ \\
  NGC -- 4 & 3 & 0.05; 0.04; 0.13 &  300; 110; 24 &               10$^8$ \\
  NGC -- 5 & 3 & 0.05; 0.04; 0.13 &  300; 110; 24 &  1.8 $\times$ 10$^8$ \\
  NGC -- 6 & 3 & 0.01; 0.04; 0.13 & 1700; 110; 24 &               10$^8$ \\
  \hline
 \end{tabular}
\end{centering}
\end{table*}

In the second set of models, we adopt the prescriptions of A05
described in Sect. 2, namely 3 episodes of SF.  The first happened
between 600 and 300 Myr ago at a rate of 0.05 M$_\odot$ yr$^{-1}$.
This episode is followed by a period of inactivity of 150 Myr and then
by a second episode lasting 110 Myr at a SF rate of 0.04 M$_\odot$
yr$^{-1}$.  After a short quiescent period (3 Myr) the last episode of
SF started.  The onset of this episode is therefore 37 Myr ago,
lasting until 13 Myr ago (24 Myr of duration in total, consistent with
the estimates of Anders et al.  2004) at a rate of 0.13 M$_\odot$
yr$^{-1}$.  Since in this case the SF rate is no longer a free
parameter, we wish to explore the effect of a different initial ISM
distribution.  Since there are still uncertainties about the total gas
mass of NGC1569 (see e.g. Stil \& Israel 2002; M\"uhle et al.  2003)
one has the freedom to test different values of the initial mass and
to see which one is more appropriate to reproduce the characteristics
of NGC1569.  In particular, we consider a ``light'' model (model NGC
-- 4), in which the total galactic \hi mass at the beginning of the
simulation is $\sim$ 10$^8$ M$_\odot$ and a model with a factor of
$\sim$ 2 more gas initially present inside the galaxy (model NGC --
5).  Finally, we consider a model in which a longer first episode of
SF is considered.  This episode lasts 1.7 Gyr, at a rate of 0.01
M$_\odot$ yr$^{-1}$.  The intermediate and the young episodes are the
same as in model NGC -- 4.  This SF history still reproduces the CMD
of NGC1569 in a satisfactory way (see A05 and Sect. 2). Table 1
summarizes the parameters adopted to model NGC1569.

\section{Chemical and dynamical evolution of NGC1569}
\label{sec:ngc1569}

\subsection{Single episode of star formation}
\label{subsec:ngc1569:single}

In this section we describe the results of the models in which a
single burst of star formation has been considered, namely NGC -- 1, 
NGC -- 2 and NGC -- 3 (see Table 1).

From the dynamical point of view, the model NGC -- 3 injects a large
amount of energy into the ISM of the galaxy.  This energy suffices to
unbind a too large fraction of the gas in a short timescale so that at
the end of the simulation more than 70 \% of the gas is lost by
galactic winds.  On the other hand, the model NGC -- 1 with a much
milder SF retains almost all the gas, showing no evidences for
galactic winds.  This disagrees with observations, since NGC1569 shows
evidences of the presence of outflowing gas (see Introduction).  For
the model NGC -- 2, a galactic wind of moderate intensity starts at t
$\sim$ 45 Myr.

\begin{figure}[ht]
\hspace{-0.5cm} \epsfxsize=9.5cm \epsfbox{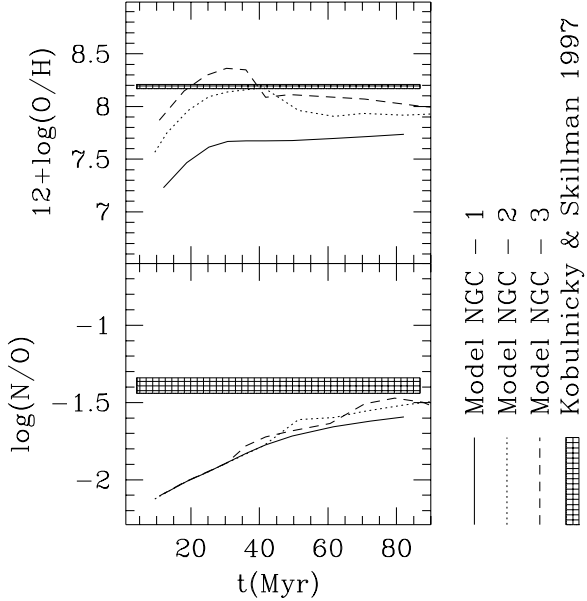} 
\caption{ Evolution of 12 + log (O/H) (upper panel) and log (N/O)
  (lower panel) for a NGC1569 model with a single episode of SF at a
  rate of 0.1 M$_\odot$ yr$^{-1}$ (model NGC -- 1; solid lines), 0.3
  M$_\odot$ yr$^{-1}$ (model NGC -- 2; dotted lines) and 0.5 M$_\odot$
  yr$^{-1}$ (model NGC -- 3; dashed lines).  The superimposed shaded
  areas indicate the observed values found by Kobulnicky \& Skillman
  (1997), with error-bars.}
\label{singbursts}
\end{figure}

In Fig.~\ref{singbursts} we plot the evolution of oxygen (upper panel)
and N/O ratio (lower panel) for these three models in comparison with
the observations of Kobulnicky \& Skillman (1997).  For the model NGC
-- 1, in spite of the fact that no oxygen is lost through galactic
winds, the final oxygen content, due to the low number of SNeII, is by
far not enough to justify the observed 12 + log(O/H).  On the other
hand, model NGC -- 3 produces a final oxygen abundance closer (but
still $\sim$ 0.2 below) to the observed value.  At the end of the
starbursting phase, the oxygen abundance is larger than the
observations, but at an age of $\sim$ 35 Myr a metal-enriched galactic
wind ensures a significant reduction of the O content of the galaxy.
In the first few tens of Myrs, the nitrogen produced by
intermediate-mass stars is not able to match the N/O ratio observed in
NGC1569.  This ratio, however, increases with time due to the delayed
production of nitrogen, reaching values close to the observations
$\sim$ 80 Myr after the onset of the burst (see Fig.~\ref{singbursts}
, lower panel).  At that time oxygen has already dropped below the
observed value.  Model NGC -- 2 shows an intermediate behaviour.  At
an age of $\sim$ 35--40 Myr it reproduces the oxygen content of
NGC1569, but at this phase of the evolution, IMS have not
significantly contributed to the global chemical enrichment of the
galaxy, therefore the N/O ratio is severely underestimated.  After the
onset of the galactic wind (t $\sim$ 45 Myr), the oxygen content
begins to decrease.

The N/O ratio of model NGC -- 2 and NGC -- 3 after the onset of the
galactic wind becomes larger than the value predicted for model NGC --
1 (unaffected by selective loss of metals).  This implies that, at the
beginning, the outflowing gas is enriched in elements produced by
SNeII (in particular $\alpha$-elements), in agreement with the
observations of Martin et al. (2002; see also below in this section).

In summary, fine-tuning the SF rate in order to fit at the same time
the oxygen abundance and the gas content of NGC1569 is not easy, since
a low SF rate can be insufficient to match the observed O/H and a
higher SF rate can drive a too powerful galactic wind, able to unbind
a too large fraction of gas initially present inside the galaxy.

It is worth reminding that the yields of MM02 might underestimate the
primary nitrogen production from IMS.  Although, as pointed out in
Sect. 3.2, this effect might be compensated by the fact that the
adopted metallicity of the stellar populations (Z=0.004) should be
considered as an upper limit, it is worth briefly discussing the
effect of a different choice of yields, in particular from IMS.
Although the set-up is different, our simulations of the evolution of
IZw18 (Recchi et al. 2004) give us a hint about the differences in N/O
produced by different sets of IMS yields.  In particular comparing
fig. 7 and fig. 10 of Recchi et al. (2004), we obtain that the
calculations of van den Hoek \& Groenewegen (1997) (together with the
Woosley \& Weaver (1995) calculations for massive stars) would produce
a final N/O $\sim$ 0.2--0.3 dex larger than the one predicted by using
the MM02 models.  A similar result has been obtained by Chiappini et
al. (2003b) by simulating the chemical evolution of the Milky Way and
of the BCD galaxies.  Looking at Fig.~\ref{singbursts} we notice that
a different choice of IMS yields would therefore produce, at the end
of the simulation ($\sim$ 80 Myr after the onset of the starburst), a
N/O consistent with the observed values of Kobulnicky \& Skillman
(1997).  However, at this age, the oxygen content of the galaxy would
be underestimated for any of the considered models.  Since the yields
of oxygen are more reliable than those of nitrogen, from the upper
panel of Fig.~\ref{singbursts} one should argue that the only
acceptable age of the burst is $\sim$ 35--40 Myr, provided that the SF
rate is larger than 0.3 M$_\odot$ yr$^{-1}$.  At this age, the
predicted N/O is $\sim$ 0.4--0.5 dex below the observed values.
Therefore, also the yields of van den Hoek \& Groenewegen would not
match the observations.  It is thus absolutely reasonable to implement
the MM02 yields.

\subsection{The X-ray emitting gas}
\label{subsec:ngc1569:xray}

\begin{figure}
 \epsfxsize=9cm \epsfbox{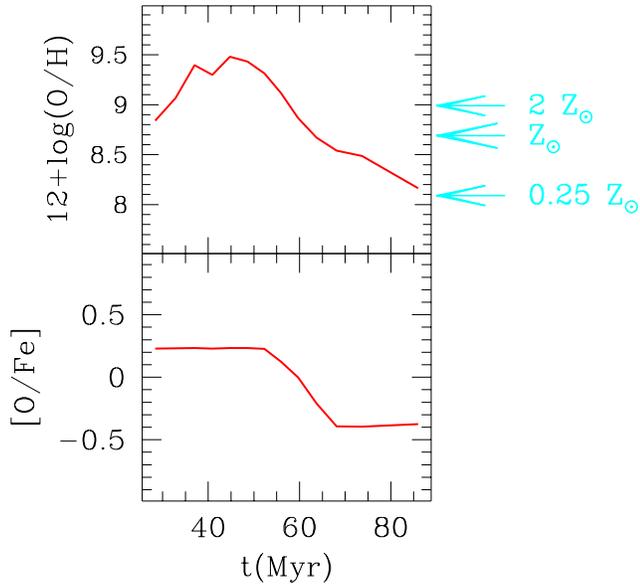} 
\caption{ Evolution of 12 + log (O/H) (upper panel) and [O/Fe]
  abundance ratio (lower panel) for the hot gas for the model NGC --
  3.  The arrows indicate the oxygen abundance of the galactic wind
  inferred by Martin et al. (2002).}
\label{short_hot}
\end{figure}

As anticipated in the introduction, Martin et al. (2002) were able to
estimate the metallicity of the hot X-ray emitting gas in the galactic
wind.  This information completes the puzzle of understanding the
metal enrichment.  Even if the single-burst models are not able to
account for the chemical and morphological properties of NGC1569, it
is nonetheless interesting to calculate the metallicity of the hot gas
(i.e. of the gas with temperatures larger than 0.3 keV) and to compare
it with the derived values of Martin et al.  (2002).  This comparison
is shown in Fig.~\ref{short_hot} for the NGC -- 3 model.  At the
moment of the onset of a galactic wind, the oxygen abundance of the
hot phase is already solar.  The arrows drawn in the plot represent
the estimated oxygen content of the galactic wind of NGC1569 (best
fit; upper and lower limits).  The oxygen abundance of the hot gas
increases continuously in this phase, since after 50 Myr massive stars
are still exploding and releasing oxygen into the interstellar medium.
At later times, however, the oxygen composition begins to decrease due
to the larger fraction of pristine gas ablated from the supershell and
entrained in the galactic wind and due to the lower amount of gas
heated up.  The [O/Fe] ratio is initially larger than solar due to the
fact that the break-out occurs when SNeIa are not yet releasing their
energy and metals into the ISM.  At later times, since the SNeIa expel
their products from the galaxy very easily (Recchi et al. 2001), the
[O/Fe] ratio decreases.  The observations of Martin et al. (2002)
however point towards a galactic wind dominated by $\alpha$-elements,
therefore an outflow probably is still triggered by SNeII.

\subsection{Three episodes of star formation}
\label{subsec:ngc1569:3}

As shown in the previous section, single SF bursts of short duration
are not able to account for the global properties of NGC1569.
According to observationally derived SF we therefore describe the
evolution of models with the SF as a gasping process, occurring in 3
different episodes as explained in Sect. 3.2.

\begin{figure}[ht]
\hspace{-0.3cm}\epsfxsize=9.5cm \epsfbox{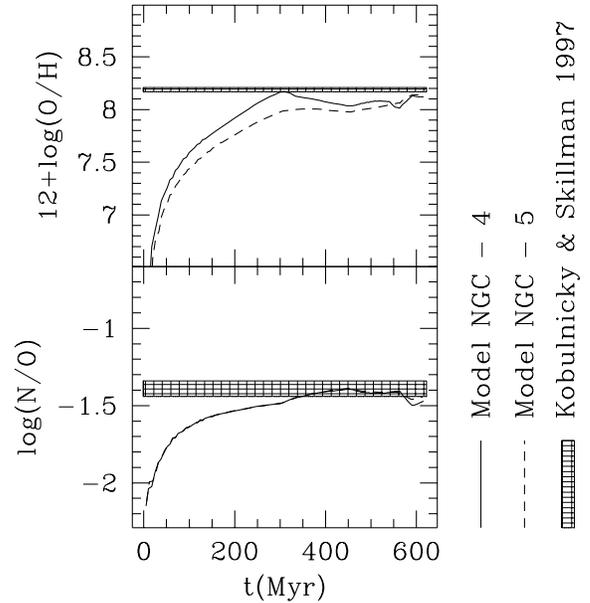} 
 \caption{ Evolution of 12 + log (O/H) (upper panel) and log (N/O)
   (lower panel) for two NGC1569 models in which the A05 SF history is
   implemented: model NGC -- 4 (solid line) and NGC -- 5 (dashed
   line).  The superimposed dashed areas represent the observed values
   found for NGC1569 (Kobulnicky \& Skillman 1997).}
\label{treb}
\end{figure}

In Fig.~\ref{treb} we show the evolution of oxygen (upper panel) and
log (N/O) (lower panel) for the models NGC -- 4 (solid lines) and NGC
-- 5 (dashed lines) and we compare them with the abundances derived by
Kobulnicky \& Skillman (1997) (dashed areas).  At the end of the
simulations (after $\sim$ 600 Myr), the oxygen is reproduced nicely by
the model NGC -- 5 and also model NGC -- 4 is close to the observed
value.  It is worth noticing that the outflow created by the
pressurized gas is very weak in the NGC -- 5 model and of moderate
intensity for the model NGC -- 4.  The fraction of oxygen lost through
the galactic wind is larger in the light model (the one with lower
total mass).  In the first hundreds of Myrs the oxygen abundance
predicted by the model NGC -- 4 is larger, since it is diluted by a
smaller amount of hydrogen.  At later times, however, the O abundance
predicted by this model slightly decreases with time, since some
oxygen is expelled from the galaxy.

The final log (N/O) predicted by these models are $-$~1.47 (model NGC
-- 4) and $-$~1.44 (model NGC -- 5).  These values slightly
underestimate the observations of Kobulnicky \& Skillman (1997), but
are still reasonably close to it, considering the observational errors
(0.05 dex).  

\begin{figure}[ht]
 \begin{center}
 \psfig{file=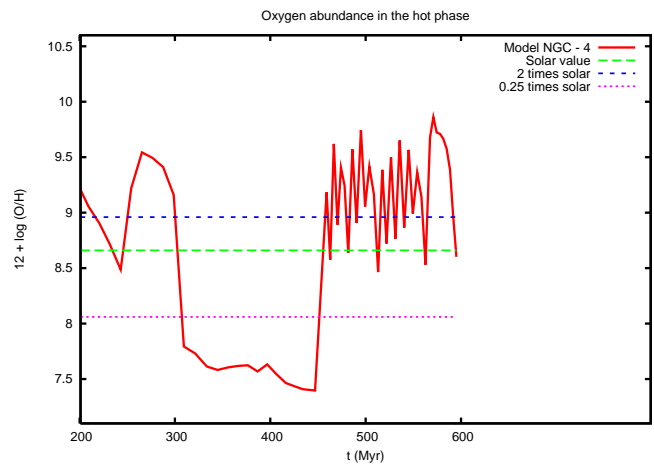,width=6.2cm,angle=270} 
\caption{ Evolution of 12 + log (O/H) for the hot gas for the model
  NGC -- 4 (see Table 1).  The oxygen abundance of the hot phase
  inferred by Martin et al. (2002) is indicated by 3 horizontal lines:
  dotted line (Z=0.25 Z$_\odot$), long-dashed line (Z=1 Z$_\odot$) and
  short-dashed line (Z=2 Z$_\odot$). }
\label{long_hot} % for cross-references 
\end{center}
\end{figure}

For model NGC -- 4 (the light model), we also have plotted the oxygen
content of the hot phase and we have compared it with the values
inferred by Martin et al. (2002).  This comparison is shown in
Fig.~\ref{long_hot}.  As one can see, the power generated by the
bursts of SF is able to heat a significant amount of oxygen, whereas,
during the quiescent periods, the fraction of hot gas drops
considerably.  Due to the poor spatial resolution of our simulation at
large radii, the oxygen abundance of the hot phase shows large
oscillations.  The average value of O/H during the last 150 Myr is
slightly larger than 2 Z$_\odot$, whereas, at the end of the
simulation, this value drops to $\sim$ Z$_\odot$ which is the best fit
of Martin et al. (2002).  However, due to the strong variation of O/H
with time, we cannot consider it as a firm prediction of our model.

\begin{figure}[ht]
 \begin{center}
 \psfig{file=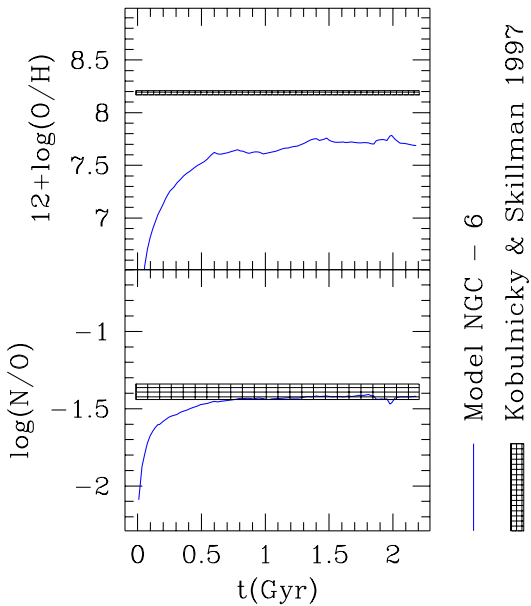,width=9.5cm} 
\caption{ Evolution of 12 + log (O/H) and log (N/O) for the model
  NGC -- 6 (see Table 1).}
\label{mod6} % for cross-references 
\end{center}
\end{figure}

We describe now the evolution of the model NGC -- 6, namely a model in
which the first episode of SF lasts 1.7 Gyr, at a rate of 0.01
M$_\odot$ yr$^{-1}$ and the other two episodes are identical to the
ones adopted in models NGC -- 4 and NGC -- 5.  The total energy
released in this episode of SF is slightly larger than in the previous
cases.  Moreover, the continuous input of energy affects the gas
distribution in front of the bubble along the $z$ direction and,
although the energy input rate is pretty mild, at $\sim$ 600 Myr a
galactic wind arises and is sustained until the end of the simulation.
This also produces a significant metal loss and, at the end of the
simulation, the oxygen content of the galaxy is $\sim$ 0.4--0.5 dex
below the observations (see Fig.~\ref{mod6}).  This model is therefore
less suited than models NGC -- 4 and NGC -- 5 to explain the observed
chemical composition of NGC1569.

Finally, we briefly mention the effect of a different IMF choice.  We
ran a model in which the IMF slope is flatter than Salpeter (x=0.85
instead of x=1.35).  As described in A05, a flatter IMF is still
marginally consistent with the observed CMD diagram of NGC1569.
Although the Salpeter IMF is able to nicely reproduce the
mass/luminosity ratio in NGC1569 (Sternberg 1998) it is worth
analyzing the results of a different IMF slope.  As expected, a model
with a flatter IMF produces a very large fraction of massive stars.
The overproduction of oxygen is partially compensated by the fact
that, due to the larger energy input from SNeII, the galactic wind is
more intense and therefore the fraction of ejected O is larger.  At
the end of the simulation the oxygen is overestimated by $\sim$ 0.2
dex compared to observations.  The nitrogen is underproduced (due to
the minor fraction of IMS) and the final log (N/O) is 0.4-0.5 dex
below the observations.  As shown in Recchi et al. (2004), a steeper
IMF slope would underproduce oxygen and overestimate log (N/O).
Therefore, the choice of a Salpeter slope seems to be the safest one
also from a chemodynamical point of view.

\subsection{The cooling of freshly produced metals}

\begin{figure*}[ht]
 \begin{center}
 \vspace{-1cm}
 \psfig{file=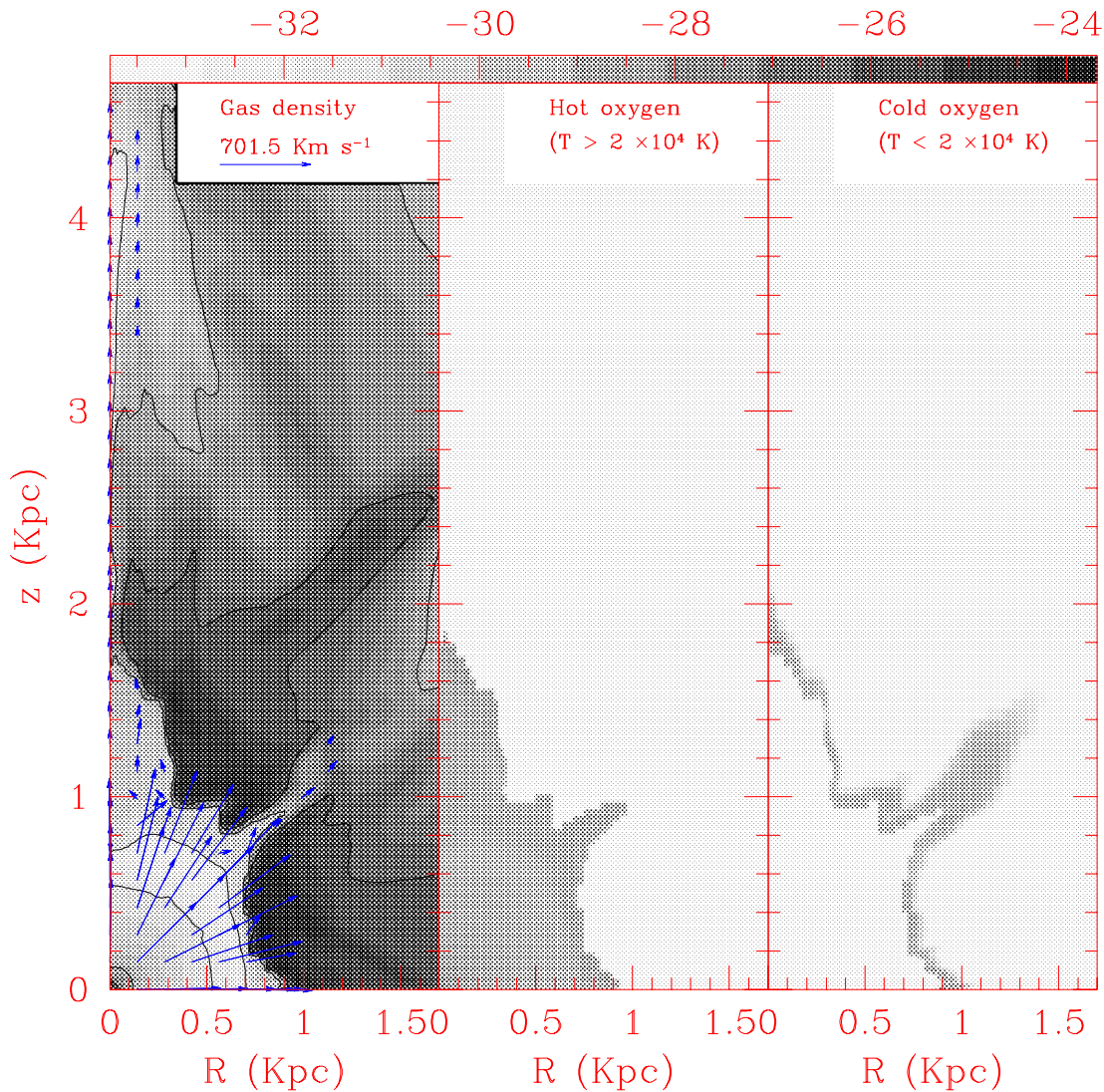,width=16cm}
\caption{Density contours and velocity fields for model NGC -- 4
  $\sim$ 15 Myr after the onset of the last episode of SF (left
  panel).  Also plot are the distribution of cold (T $<$ 2 $\times$
  10$^4$ K) and hot (T $\geq$ 2 $\times$ 10$^4$ K) oxygen at the same
  evolutionary time (left and central panel, respectively).  }
\label{ox_ngc4} % for cross-references 
\end{center}
\end{figure*}

As one can see in Fig.~\ref{treb}. the effect of the last, more
intense burst of SF (occurring at an evolutionary time of $\sim$ 560
Myr) is almost unrecognizable in the NGC -- 5 model and also in the
NGC -- 4 model a very mild increase in oxygen by $\sim$ 0.1 dex (and a
relative drop in N/O) is visible.  This is due to the fact that the
hot cavity carved by the previous episodes of SF is very large and a
significant fraction of the freshly produced gas is already carried
outside the galaxy.  Therefore, the freshly produced metals are
released in a hot and rarefied environment and their cooling
time-scales are very long.  To visualize this effect, we consider the
evolution of model NGC -- 4 $\sim$ 15 Myr after the onset of the last
burst of SF.  In Fig.~\ref{ox_ngc4} we plot the density contour and
velocity field (left panel) of this model, together with the
distribution of cold (T $<$ 2 $\times$ 10$^4$ K) and hot (T $\geq$ 2
$\times$ 10$^4$ K) oxygen (left and central panel, respectively).  It
is evident that only a negligible fraction of freshly produced oxygen
has cooled down.  Most of the cold oxygen is located in a tongue-like
structure at $R$ $\sim$ 1 kpc, $z$ $\sim$ 1 kpc that has been produced
by Kelvin-Helmholtz instabilities resulting from the shear of the hot
gas when flowing through the funnel visible at $R$ $\sim$ 0.7 kpc, $z$
$\sim$ 0.8 kpc in the density distribution plot (left panel).  The
bulk of the freshly produced oxygen is in a hot phase (central panel)
and will stay hot until the end of the simulation.

\begin{figure*}[ht]
 \begin{center}
 \vspace{-4cm}
 \psfig{file=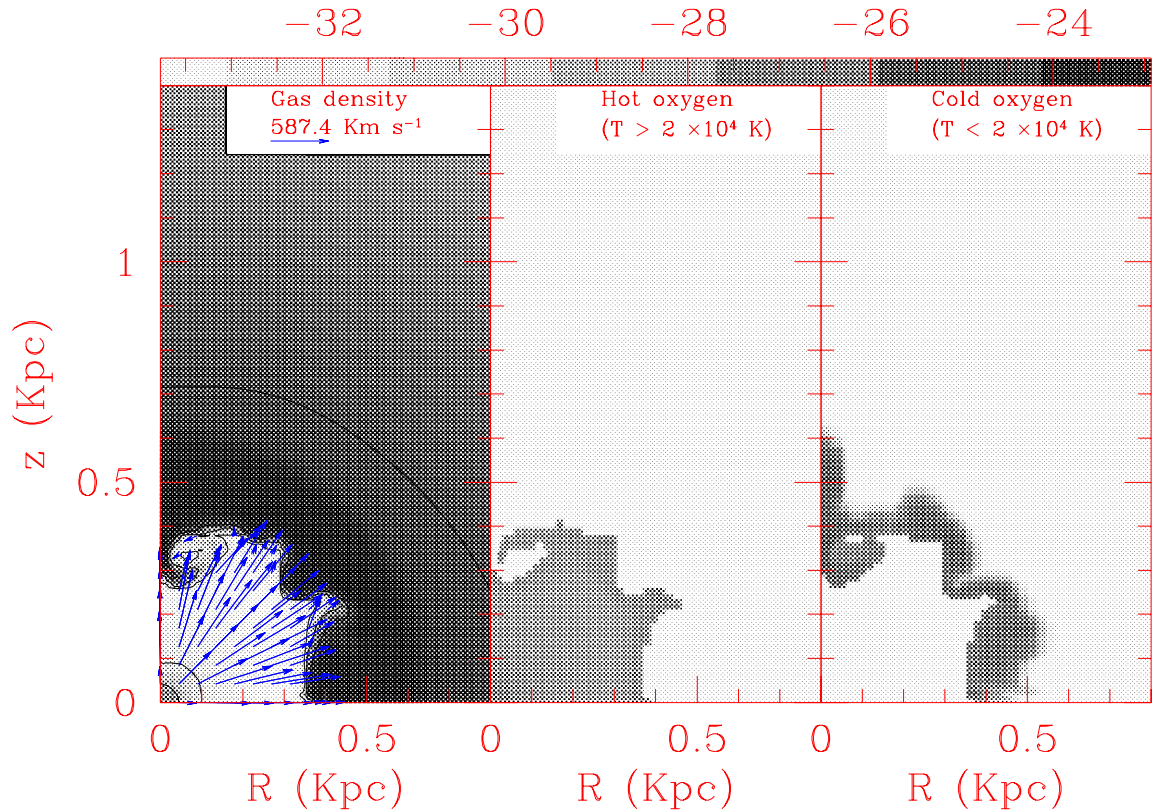,width=16cm}
\caption{As Fig.~\ref{ox_ngc4} but $\sim$ 80 Myr after the onset of 
the first episode of SF.  }
\label{ox_1b} % for cross-references 
\end{center}
\end{figure*}

The fate of the oxygen produced in the previous episodes of SF is
significantly different.  To demonstrate that, in Fig.~\ref{ox_1b} the
same plot (density distribution, distribution of cold and hot oxygen)
is shown, but for the oxygen produced by the first episode of SF by
the model NGC -- 4 after an evolutionary time of $\sim$ 80 Myr.  As
one can see, in this case the evolution of the superbubble is still
relatively slow (due to the low SF rate) and a significant fraction of
the freshly produced oxygen can be pushed close to the cavity walls,
where eddies (particularly evident is an eddy at $R$ $\sim$ 0.2 kpc,
$z$ $\sim$ 0.35 kpc), thermal instabilities and thermal conduction are
able to broaden the interface between the hot cavity and the cold
shell, resulting in a relatively fast cooling of a significant amount
of freshly produced metals.  The fact that the chemical elements
produced by the last episode of SF do not cool down (and therefore do
not contribute to the chemical composition of the \hii regions) helps
to explain why most of the \hii regions in NGC1569 show similar
chemical compositions (Kobulnicky \& Skillman 1997).  This composition
is in fact mostly due to the metal production of the older episodes of
SF, which have had enough time to cool down and to disperse throughout
the galaxy.

\subsection{The Intergalactic Medium enrichment}

It is now well established that the iron content in the Intracluster
Medium (ICM) is pretty large, ranging from $\sim$ 1/5 to about half
solar, with a mean value around 1/3.  The [$\alpha$/Fe] ratios are
close to solar (Renzini 1997), or even undersolar ([O/Fe] $\sim$ -0.3)
for M87 (Gastaldello \& Molendi 2002; Matsushita; Finoguenov \&
B\"ohringer 2003).  Some dependence on the cluster metallicity or
temperature may also be present (Finoguenov, David \& Ponman 2000;
Loewenstein 2004).  We thus need to invoke a mechanism able to enrich
iron in the ICM to a level at least comparable to the enrichment of
$\alpha$-elements.

A first assumption is that Pop. III stars (Ostriker \& Gnedin 1996) or
proto-galaxies at very high z (Mori, Ferrara \& Madau 2002) can
increase the metallicity of the ICM to the levels observed nowadays.
However, observations have established that the enrichment of the ICM
at $z$ $\sim$ 3 is very low, about two orders of magnitude lower than
the metal content reached at $z$ $\sim$ 0 (see Aguirre et al. 2001 and
references therein).  Most of the ICM enrichment should have occurred
between z $\sim$ 2 and z $\sim$ 1, since there is no significant
evolution of the Fe metallicity out to z $\sim$ 1 (Tozzi et al. 2003).
Thus, these pre-galactic object can enrich the ICM but only to a very
low level so that most of the ICM enrichment has to come from
galaxies.

We can only imagine two processes able to enrich the ICM: ram pressure
stripping and galactic winds.  Ram pressure stripping may be very
effective in removing the gas in dwarf galaxies (Mori \& Burkert 2000;
Marcolini, Brighenti \& D'Ercole 2003), and its effect can be
significant also in larger disk galaxies (Roediger \& Hensler 2005).
However, cosmological simulations by Aguirre et al. (2001) show that
ram pressure stripping can only explain $\sim$ 1/12 of the metals
observed in the ICM.  Moreover, ram pressure stripping affects mostly
the external part of the discs (Roediger \& Hensler 2005), thus
involving regions in which the iron enrichment might be negligible.
Galactic winds instead are able to reproduce both the metallicity and
the thermal history of the ICM (Pipino et al. 2002).

In the literature it is often claimed that the SNII products (thus in
particular $\alpha$-elements) should dominate the gas ejected through
galactic winds (Bradamante, Matteucci \& D'Ercole 1998; D'Ercole \&
Brighenti 1999, among others).  The observation of the outflow
metallicity in NGC1569 seems to confirm this conclusion.  The same
happens also in our models, since, as we have seen in
Fig.~\ref{short_hot} and in Fig.~\ref{long_hot}, during and
immediately after the bursts of SF a very large amount of oxygen is
transported out of the galaxy through the galactic wind.  As time goes
on, in the models in which a galactic wind is formed, the wind is
progressively enriched in iron-peak elements, due to the minor work
required to escape the galactic potential well once a funnel has been
carved (see also Recchi et al. 2001; Recchi et al. 2004).  This can
produce a significant release of iron from dwarf galaxies to the ICM,
which stops only when the energy provided by SNeIa is no longer able to
sustain the galactic wind and the center of the galaxy is refilled
with cold gas.  This happens a few hundreds Myrs after the end of the
SF activity (Recchi \& Hensler, in preparation).

It is however worth pointing out that, although the number of dwarf
galaxies in clusters is very large, the cluster luminosity (and mass)
is dominated by giant elliptical galaxies.  Gibson \& Matteucci (1997)
demonstrated that, under conservative hypotheses, dwarf galaxies can
provide at most $\sim$ 15 \% of the ICM gas through galactic winds,
whereas the giant ellipticals are responsible for $\sim$ 20 \% and the
remaining $\sim$ 65 \% being of primordial origin.  Indeed also giant
elliptical galaxies can provide a significant iron ICM enrichment.
This depends on the interplay between the energy input from SNeIa and
the mass restored to the ISM from dying stars.  If the secular type Ia
SN rate decreases faster than the rate of mass return from IMS, the
energy of SNeIa can sustain a supersonic galactic wind (expelling a
significant fraction of iron) for several Gyr.  This wind will later
on turn to a subsonic wind and eventually an inflow (Ciotti et al.
1991; Renzini et al. 1993; Pipino et al. 2002; Pipino et al. 2005).
The reverse sequence applies in case of rates of SNeIa milder than the
rates of mass return from dying stars.  In this case early inflows are
eventually followed by galactic winds (Loewenstein \& Mathews 1987).

\subsection{The evolution of log(N/O) and the ``plateau'' problem}

Nitrogen determinations are available for a bulk of DIrrs and BCD
galaxies (see e.g. Pagel et al. 1992; Masegosa, Moles \&
Campos-Aguilar 1994; Kobulnicky \& Skillman 1996; van Zee, Haynes \&
Salzer 1997; Izotov \& Thuan 1999; V\'ilchez \& Iglesias-P\'aramo
2003; Skillman, C\^ot\'e \& Miller 2003, among others).

Nitrogen results from CNO processing of O and C during hydrogen
burning.  This can happen either from O and C originally present in
the star (secondary N production), or from O and C produced {\it in
situ} and dredged up to the H-burning layer.

The behaviour of the N/O ratio as a function of O/H has provided the
main observational constraint to the processes for N formation
(secondary and/or primary).  Assuming instantaneous recycling (i.e.
assuming that all the stars with masses larger than 1 M$_\odot$ die
instantaneously), if N is produced in a secondary way, it increases
proportionally to the metal content of the galaxy and the track in the
N/O vs. O/H diagram is a straight line at a 45$^{\rm o}$ slope.  If
instead the nitrogen is produced in a primary way, it evolves
independently of the O content of the galaxy (Tinsley 1980; Matteucci
1996).  However, the effect of the delayed production of N, due to its
progenitors, must also be taken into account.  Observations of
metal-poor (log (O/H) $\simlt$ $-$ 4) \hii regions in dwarf galaxies
(van Zee et al. 1997; Izotov \& Thuan 1999) have revealed that N/O is
almost independent of O/H and this is what is expected if N is
produced in a primary way.  On the other hand, for high metallicity
\hii regions, N/O increases almost linearly with O, indicating that
most of N is produced in a secondary way.  A large scatter ($\pm$ 0.3
dex) in the N/O vs. O/H diagram is present at large metallicities, and
this has been interpreted as the consequence of the delayed production
of N, resulting in a ``saw-tooth'' evolution in this diagram (Pilyugin
1993; Marconi, Matteucci \& Tosi 1994; Olofsson 1995).  At lower O the
scatter identified by Izotov \& Thuan (1999) is very small.  The
conclusion of these authors is that these metal-poor galaxies are
undergoing their first burst of SF and the observed N comes only from
primary production in massive stars.  Observations of DLAs (see e.g.
C\'enturion et al. 2003 and references therein) show a significant
population of objects with N/O lying at 0.7 -- 0.8 dex below the
plateau indicated by Izotov \& Thuan (1999).  If N is produced in a
primary way in massive stars, as invoked for metal-poor BCDs, it would
be hard to explain this population of objects with very low N content
(but see also Izotov, Schaerer \& Charbonnel 2001).  Moreover, none of
the nucleosynthetic prescriptions available in literature are able to
reproduce the observed N/O ratio by means of massive stars alone
(Recchi et al.  2002).

\begin{figure*}[ht]
 \begin{center}
 \psfig{file=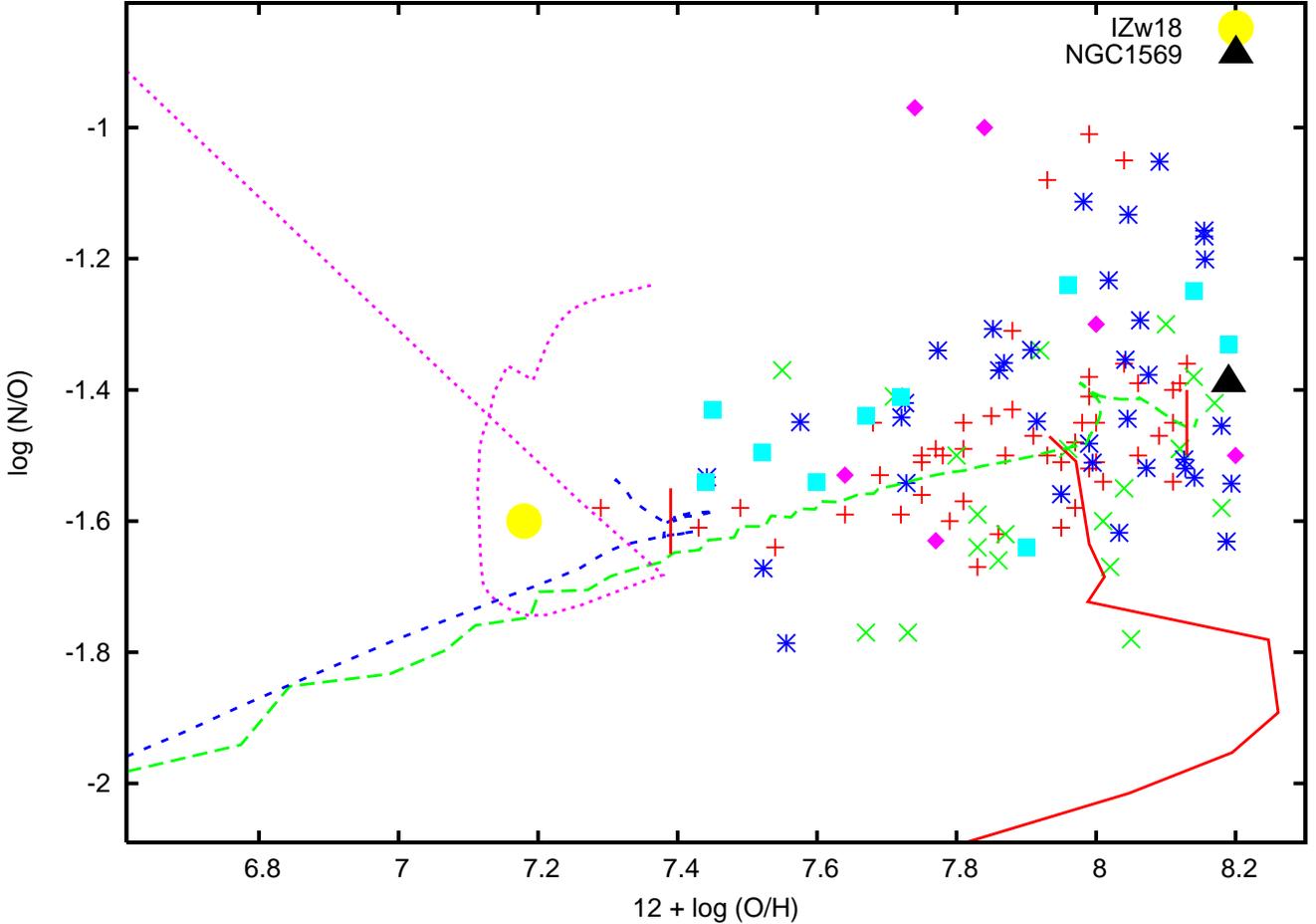,width=12.7cm,angle=270}
\caption{ log (N/O) vs. 12 + log (O/H) in metal-poor BCD and DIrr
  galaxies.  Filled circle is the IZw18 value calculated by Izotov \&
  Thuan (1999), whereas the filled triangle is the NGC1569 value from
  Kobulnicky \& Skillman (1997).  Pluses are other galaxies detected
  by Izotov \& Thuan (1999).  The collection of data by van Zee et al.
  (1997) is shown by crosses.  The measurements of BCD galaxies by
  Kobulnicky \& Skillman (1996) are indicated by asterisks.  Filled
  rhombs represent the values tabulated by V\'ilchez \&
  Iglesias-P\'aramo (2003).  The other data points (filled squares)
  are taken from different sources.  Also shown is the evolution in
  the N/O vs. O/H plane of two NGC1569 models: model NGC -- 3 (solid
  line) and NGC -- 4 (long-dashed line).  For comparison, we also plot
  two model reproducing the evolution of IZw18 after 2 instantaneous
  bursts (dotted line) and after a gasping star formation derived by
  Aloisi et al. (1999) (short-dashed line).  These last two models are
  taken from Recchi et al. (2002) and Recchi et al. (2004),
  respectively.  A vertical line indicates, for model NGC -- 4 and for
  the gasping IZw18 model the onset of the last burst of SF.}
\label{obs} % for cross-references 
\end{center}
\end{figure*}

We have collected the most recent data available in literature about O
and N abundances in very metal-poor galaxies (with 12 + log (O/H)
$\leq$ 8.2).  In Fig.~\ref{obs} we plot these data, together with the
evolution in the N/O vs. O/H plane of the bursting NGC -- 3 (solid
line) and the gasping NGC -- 4 (long-dashed line) models.  For
comparison, we also plot two models describing the evolution of IZw18,
following two instantaneous bursts of SF (dotted line) and a gasping
SF history, with a weak episode of SF lasting 270 Myr, a quiescent
period of 10 Myr and a more intense burst lasting 5 Myr (short-dashed
line).  These models have been comprehensively described in Recchi et
al. (2002) and Recchi et al. (2004), respectively.  In the bursting
model (dotted line), only the evolution after the onset of the second
burst is plotted.  In this sample of data, we can notice the presence
of a considerable spread, at variance with the results of Izotov \&
Thuan (1999).  Moreover, Kennicutt \& Skillman (2001) have pointed out
that the observed narrow dispersion in the Izotov \& Thuan (1999)
observations itself is difficult to understand given the associated
uncertainties in reddening corrections, the ionization correction
factor, and the estimated temperature in the O$^+$ zone.  The
``plateau'' at log (N/O) $\simeq$ $-1.6$ needs therefore more
statistics in order to be firmly established, also considering that,
in order to have a homogeneous sample of objects, in this figure we
have not plotted the N/O of DLA systems, whose inclusion would arise
in a considerable increase of the scatter.  It is however worth
pointing out that Henry, Edmunds \& K\"oppen (2000) performed
simulations in which they were able to fit the trend of N/O for BCDs
with a very low SF rate, whereas a higher SF rate were invoked to
explain the DLAs with low oxygen and low N/O.

The comparison between the observed data and the evolutionary paths
allows us to understand how the N/O ratio changes in the first phases
of the evolution of the galaxy.  However, we have to bear in mind
that, as shown by Chiappini et al. (2003a), the observed data in the
N/O vs. O/H diagram do not give us any information about time
evolution.  They are simply the present-day status of each single
galaxy.  As we have seen in the previous sections, model NGC -- 3
(single-burst model) is unable to account for the O/H and N/O ratio at
the same time, whereas model NGC -- 4 gets very close to the
observations of Kobulnicky \& Skillman (1997).  A vertical line
indicates also the moment of the onset of the last episode of SF.
This happens when the 12 + log (O/H) has already reached 8.13 and
little or no variations of the chemical composition of the galaxy
happen afterwards.  The last burst of SF is therefore unable to modify
the chemical composition of the galaxy observed in the \hii region,
due to the fact that the metals produced by the last SF episode are
injected in a too hot phase and cannot contribute to the total
chemical budget of the galaxy (see Sect. 4.4).  Models describing
bursting and gasping SF, for both IZw18 and NGC1569, show similar
patterns: a bursting model produces huge variations of the chemical
composition of the galaxy in short time-scales, whereas the gasping
model predicts a mild increase of oxygen and of N/O and a
stabilization of the chemical composition at later times (also for the
gasping IZw18 model we have plotted a vertical line at the onset of
the last burst of SF).  It is also visually clear from this plot that
gasping models are much more adequate to explain the similar N/O
ratios in metal-poor galaxies.  If the plateau suggested by Izotov \&
Thuan (1999) is real, we thus have to invoke the scenario of weak and
continuous episodes of SF in order to explain the observations.
Episodes of SF older (and weaker) than the ones considered in our
simulations are not ruled out by A05.  When considering episodes of SF
older than $\sim$ 2 Gyr, their nature (bursting or gasping) become
less relevant, since the abundance ratios are dominated by the SF
occurring in the last $\sim$ 2 Gyr (Romano, Tosi \& Matteucci 2005b).
Finally, in spite of a difference in the oxygen abundance of $\sim$ 1
dex between IZw18 and NGC1569, the difference in (N/O) is $\sim$ 0.2,
which might again indicate a mild nitrogen production in some
long-lasting episode of star formation, as it happens in the model NGC
-- 4 (long-dashed line in Fig.~\ref{obs}).

\subsection{Models with a big infalling cloud}
\label{subsec:ngc1569:bigclo}

\begin{figure*}[ht]
 \begin{center}
 \vspace{-3.5cm}
 \psfig{file=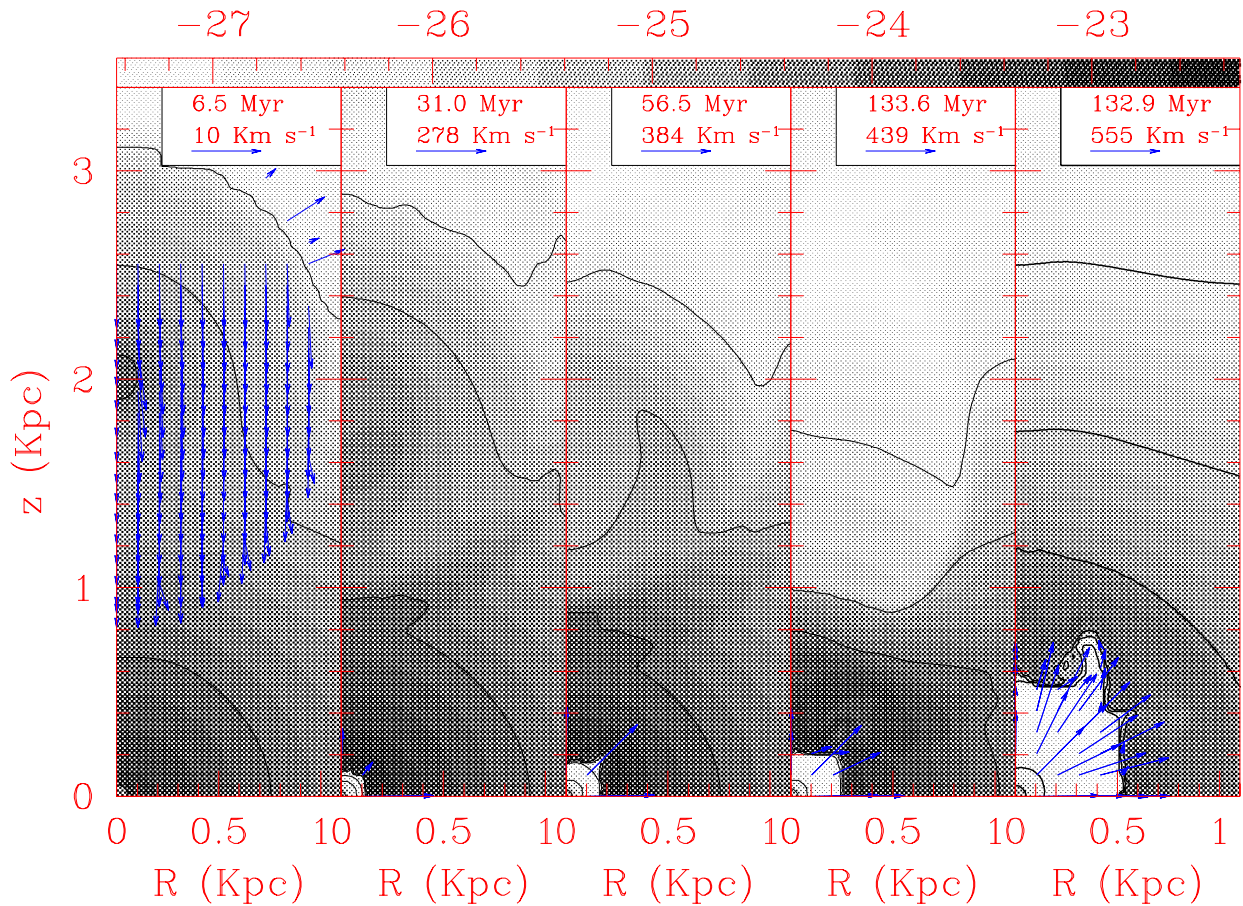,width=17cm}
\caption{Density contours and velocity fields for model NGC -- 4+BC 
  (heavy cloud) at four different epochs (evolutionary times are
  labelled in the box on top of each panel).  In the last panel the
  density contours and velocity fields for the reference model NGC --
  4 at a similar evolutionary time as the last plot of model NGC --
  4+BC are shown.  The logarithmic density scale is given in the strip
  on top of the figure.  In order to avoid confusion, velocities with
  values lower than 1/10 of the maximum value (indicated for each
  panel in the upper right box) are not drawn.  }
\label{snap1} % for cross-references 
\end{center}
\end{figure*}

As described in the introduction, extended \hi clouds surround NGC1569
(Stil \& Israel 1998).  In particular, a series of gas clumps in the
southern part of the galaxy, probably connected with the \hi arm, led
M\"uhle et al. (2005) to the hypothesis that these clumps can be
attributed to the debris of a tidally disrupted big cloud infalling
towards NGC1569.  The mass of this complex is difficult to assess.  The
lower limit given by M\"uhle et al. (2005) is 1.2 $\times$ 10$^7$
M$_\odot$ (the sum of the mass of all the detected groups of clouds),
but some \hi could have been already accreted.  If this complex is
similar to the HVCs in the Local Group, one has to expect masses
larger than a few 10$^7$ M$_\odot$ (Blitz et al. 1999; de Heij, Braun
\& Burton 2002).  From a chemical point of view, the expected effect
of an infalling cloud is to dilute the gas in the galaxy, allowing for
a reduction of the metallicity without altering the abundance ratios
(K\"oppen \& Hensler 2005).  In reality, the coexistence of hot and
cold gas leads to interfaces due to heat conduction where from
analytical estimates under ``normal'' conditions cold gas should
suffer evaporation.  From a dynamical point of view, this leads to
mass-loaded flows (Dyson \& Hartquist 1987) in which the specific
energy is reduced so that an outflow can be hampered.  More realistic
hydrodynamical simulations of self-gravitating interstellar clouds
could, however, show that their disruption is prevented (Vieser \&
Hensler 2005).  Moreover, an infalling cloud produces ram pressure
which acts like a ``cap'' for the expansion of the superbubble.

In order to study the effect of a big cloud infalling towards the
galaxy, we assume a mass of 2 $\times$ 10$^7$ M$_\odot$. The initial
position of the cloud is 2 kpc away from the center of the galaxy
along the polar axis (due to the assumed axial symmetry of the system,
this is the only reasonable initial configuration).  The chosen
density profile of the cloud is a power law, with a slope of -1.7 (de
Heij et al. 2002).  The infalling velocity of this cloud is 10 km
s$^{-1}$, similar to the local sound speed, and its radius is 1 kpc.
The metallicity of the cloud is set to 0.  Two possible SF scenarios
have been analyzed: a single burst of SF, as in model NGC -- 3,
therefore with a SF rate of 0.5 M$_\odot$ yr$^{-1}$ and a model,
analogous to model NGC -- 4 in which the implemented SF history is the
one inferred by A05.  These models are named NGC -- 3+BC and NGC --
4+BC, respectively.  For the model NGC -- 4+BC we also consider the
possibility of a lighter infalling cloud, namely a cloud with a total
mass of 10$^7$ M$_\odot$ (half of the standard value).

In Fig.~\ref{snap1} snapshots of the evolution of the model NGC --
4+BC (with the standard 2 $\times$ 10$^7$ M$_\odot$ cloud) during the
first $\sim$ 130 Myr are shown.  In the left panel, the big cloud and
its infalling motion is clearly visible.  It is also worth noticing
that we do not put the cloud in pressure equilibrium with the
surrounding ISM, therefore the cloud tends to expand towards the
regions of minor pressure.  The pressure of the infalling cloud limits
the expansion of the superbubble and shrinks it, mostly in the polar
direction.  In the second-last plot shown in Fig.~\ref{snap1} the
cavity has an extension of $\sim$ 250 $\times$ 200 pc$^2$.  For
comparison, we also plot in Fig.~\ref{snap1} (right panel) the density
contours and velocity fields for the reference model NGC -- 4 at an
evolutionary time similar to the last one considered for the model NGC
--4+BC.  As one can see, the cavity carved by the ongoing episode of
SF is much larger in this case ($\sim$ 500 $\times$ 600 pc$^2$) and
the density in front of the superbubble is also much smaller, since in
the model NGC -- 4+BC the big cloud has compressed the gas in front of
the superbubble along the $z$ direction up to densities of the order
of $\sim$ 10$^{-23}$ g cm$^{-3}$.  The effect of the infall is
therefore to hamper the development of a galactic wind and to confine
the superbubble well inside the galactic region.  Only in the model
NGC -- 4+BC with light cloud a moderate galactic wind is still present
and some gas (and metals) is lost.  Moreover, the cloud on its path
towards the galaxy sweeps up and drags some halo gas present in the
outer regions.  The final gas mass inside the galaxy is therefore
larger than the initial one.

Also the model NGC -- 3+BC, in spite of the large SF rate, does not
experience large outflows and the mass of the \hi gas increases with
time.  It is worth noticing that, as we have pointed out in the
previous sections, NGC1569 does show a prominent outflow (Martin et
al. 2002), at variance with the results of these simulations.
Therefore, we either have to consider a larger input of energy into
the system, or have to consider a different infall direction of the
cloud.  Indeed, observations show that this \hi complex seems to wrap
around the disk of NGC1569 and to approach the galaxy from the western
side (M\"uhle et al. 2005).  Such an infall geometry is however
impossible to reproduce with a 2-D simulation in cylindrical
coordinates.  Although the 2-D symmetry can only account for a
vertical infall along the $z$-axis, one can already learn from this
kind of simulations to what extent the pressure of the cloud hampers
the expansion of the superbubble.  Moreover, since there are no
extensive calculations available in the literature about
chemodynamical evolution of galaxies with infall, it is extremely
interesting to study what changes does this infall produce on the
global chemical budget of the galaxy (see e.g. K\"oppen \& Hensler
2005).

\begin{figure}[t]
 \epsfxsize=9cm \epsfbox{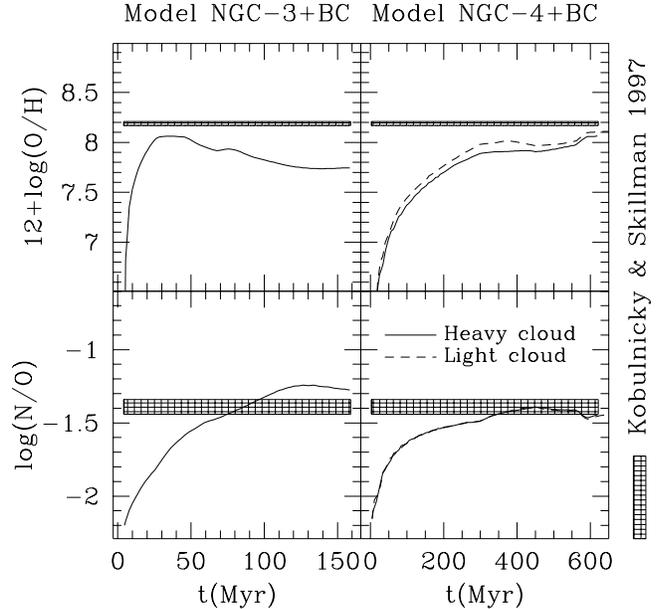} 
\caption{ Evolution of 12 + log (O/H) (upper panels) and log (N/O)
  (lower panels) for two NGC1569 models in which the infall of a big
  cloud towards the center of the galaxy is taken into consideration.
  Left panels show a model in which a single burst of SF, lasting 25
  Myr, is considered (model NGC -- 3+BC), whereas in the right panels
  the evolution of a model implementing the SF rate inferred by A05 is
  shown (model NGC -- 4+BC).  In this case we consider two possible
  masses of the big cloud: 10$^7$ M$_\odot$ (light cloud; dashed line)
  and 2 $\times$ 10$^7$ M$_\odot$ (heavy cloud; solid line).}
\label{no_bigcloud}
\end{figure}

The evolution of oxygen and log (N/O) for these models is shown in
Fig.~\ref{no_bigcloud}.  Since almost no metals are lost through the
galactic wind, due to the hampering effect of the infalling cloud, the
oxygen produced by the various episodes of SF is mostly retained by
the galaxy and mixes with the surrounding ISM.  Only for the model NGC
-- 4+BC with light cloud (dashed lines in the right plots) a weak
galactic wind produces a drop in the oxygen content at $\sim$ 450 Myr.
In the model NGC -- 3+BC, in which the oxygen is produced only in the
first $\sim$ 50 Myr, the dilution effect due to the infalling gas is
more evident.  Part of the unprocessed gas dragged by the cloud mixes
with the metals released by the burst of SF, reducing the global
metallicity of the galaxy.  The final log (N/O) of model NGC --4+BC is
consistent with the observations, whereas this model slightly
underestimates the final oxygen content of the galaxy (by $\sim$ 0.1
-- 0.15 dex depending on the mass of the big cloud).  The oxygen
abundance in the model NGC -- 3+BC always remains more than 0.2 dex
below the observations.

\section{Conclusions}
\label{sec:conclusion}

By means of a 2-D hydrodynamical code, we have studied the chemical
and dynamical evolution of model galaxies resembling NGC1569, a
gas-rich dwarf galaxy in the aftermath of an intense burst of SF.  We
have considered either one episode of SF of short duration (bursting
SF), or more complex behaviours, in which the galaxies have
experienced long-lasting episodes of SF in the past, separated from
the last, more intense burst, by short periods of inactivity (gasping
SF).

Models with a bursting SF are generally unable to account for the
chemical and morphological properties of this object, since they
either severely underproduce O, or inject too much energy into the
system, enough to unbind a too large fraction of the gas initially
present in the galaxy.

The best way to reproduce the chemical composition of NGC1569 is
therefore assuming long-lasting, continuous episodes of SF of some
hundreds Myrs of age and a recent and more intense short burst.
Adopting the SF prescriptions derived from the comparison of the
color-magnitude diagrams with synthetic ones (A05) we produce results
in good agreement with the observations, if the yields of MM02 are
implemented.  A simulation in which a big cloud is falling towards the
center of the galaxy along the polar axis produces also results in
good agreement with the observed chemical composition of NGC1569, but
the cloud inhibits almost completely the formation of a galactic wind,
in contradiction with observations.

In most models with gasping SF, the final chemical composition of the
galaxy reflects mostly the chemical enrichment from old stellar
populations.  In fact, if the first episodes of SF are powerful enough
to create a galactic wind or to heat up a large fraction of the gas
surrounding the star forming region, the metals produced by the last
burst of SF are released in a too hot medium to be observed or are
directly ejected from the galaxy through the wind.  They do not have
the chance to pollute the surrounding medium and contribute to the
chemical enrichment of the galaxy.  This results confirms the finding
of Martin et al. (2002) that most of the oxygen produced in the last
episode of SF is in the hot, X-ray emitting phase.

\begin{acknowledgements}
  
  S.R. acknowledges generous financial support from the Alexander von
  Humboldt Foundation and Deutsche Forschungsgemeinschaft (DFG) under
  grant HE 1487/28-1.  S.R. would also like to thank Stefanie M\"uhle
  and Donatella Romano for their suggestions and fruitful discussions,
  Monica Tosi for providing a new CMD of IZw18 in advance of
  publication and Stefan Hirche for careful reading the first version
  of the manuscript.  We finally thank the anonymous referee for
  thoughtful comments and suggestions, which have greatly improved the
  paper.
\end{acknowledgements}

\end{document}